\documentstyle[11pt,newpasp,twoside,epsf]{article}
\newcommand{\Ms}{\mbox{$\,M_{\odot}$}}
\begin{document}

\title{Common-Envelope Evolution: the Nucleosynthesis in Mergers of
Massive Stars}
\author{N. Ivanova \& Ph.\ Podsiadlowski}
\affil{Oxford University, Oxford, OX1 3RJ, UK}
\author{H. Spruit}
\affil{Max-Planck-Institut f\"{u}r Astrophysik, 
Karl-Schwarzschild-Strasse 1, 85740 Garching, Germany}

\begin{abstract}
We study the merging of massive stars inside a common envelope for
binary systems consisting of a red supergiant with a mass of
$15\,$--$\,20\,M_{\odot}$ and a main-sequence companion of
$1\,$--$\,5\, M_{\odot}$.  We are particularly interested in the stage
when the secondary, having overfilled its Roche lobe inside the common
envelope, starts to transfer mass to the core of the primary at a very
high mass-transfer rate and the subsequent nucleo\-synthesis in the
core-impact region. Using a parametrized model for the structure of
the envelope at this stage, we perform 2-dimensional
hydrodynamical calculations with the Munich Prometheus code to
calculate the dynamics of the stream emanating from the secondary and
its impact on the core of the primary. We find that, for the lower end
of the estimated mass-transfer rate, low-entropy, hydrogen-rich
material can penetrate deep into the primary core where
nucleosynthesis through the hot CNO cycle can take place and that
the associated neutron exposure may be sufficiently high for
significant s-processing.  For mass-transfer rates at the high
end of our estimated range and higher densities in the stream, the
stream impact can lead to the dredge-up of helium, but the neutron
production is too low for significant s-processing.

\end{abstract}

\section{Introduction}

The merging of two stars inside a common envelope is a common, but
very poorly understood phase of binary evolution.  It is the
consequence of dynamical mass transfer and typically occurs when the
mass-losing star is a (super-)giant expanding more rapidly as a result
of mass loss than its Roche lobe (for a general discussion see, e.g.,
Podsiadlowski 2001).  The secondary then orbits around the core of the
primary within a common envelope which is composed of material
from the giant's envelope. Due to drag
forces, the orbit of the binary slowly decays, and the envelope is
spun up as a result of the transfer of angular momentum from the
orbital motion to the envelope.  The spiral-in stage ends either when
the envelope is ejected or, as in the case considered here, when the
spiraling-in secondary starts to fill its own Roche lobe and begins to
transfer mass to the core of the primary. Eventually this leads to the
complete merger of the two stars.  We are particularly interested in
the question whether the merger of two massive stars (with masses
$M_1\simeq 15\,$--$\,20\Ms$ and $M_2\simeq 1\,$--$\,5\Ms$) is
accompanied by unusual nucleosynthesis, leading to a merger product, a
rapidly rotating supergiant, with anomalous chemical abundances (as
is, for example, observed in the progenitor of SN~1987A; Podsiadlowski
1992).

In this contribution, we give a brief outline of how we model the
merger of two massive stars.  In Sections~2 and 3 we describe how we
determine the structure of the common envelope during the spiral-in
phase and how we calculate the stream-core interaction,
respectively. In Section~4 we present some of the results of our study
to date.

\section{Common-Envelope Evolution: the Spiral-in Phase}

\begin{figure}
\plotfiddle{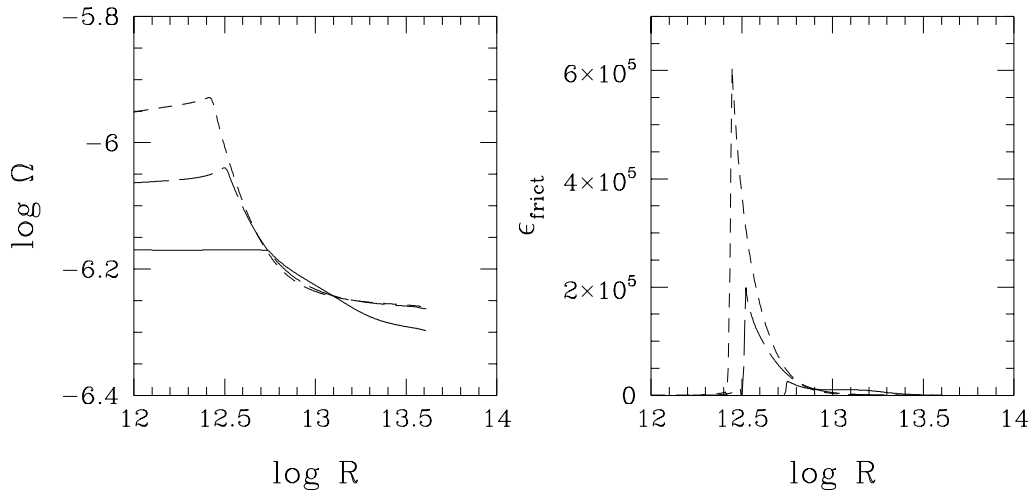}{6cm}{0}{70}{70}{-210}{-115}
\caption{Examples of the angular velocity profile and the viscous heat
        generation rate in the common envelope during the spiral-in
        stage (different curves correspond to different times and
        hence to different positions of the secondary inside the
        common envelope). All results are for a binary with a primary
        of $15\,M_{\odot}$ and a secondary of $5\, M_{\odot}$.}
\end{figure}

In order to follow the spiral-in of a star inside a common-envelope
phase after the initial rapid phase (see Podsiadlowski 2001), we
calculate the structure of the primary by simulating the effects of a
spiraling-in secondary (i.e., the additional energy sources due to
the frictional luminosity caused by the differential rotation between
the spiraling-in co-rotating binary core and the envelope, and the
accretion luminosity caused by the accretion of envelope material
onto the secondary, limited to the Eddington luminosity). We also
include the effect of the additional gravity of the secondary on the
structure of the envelope.

To follow the evolution of the secondary and determine its structure,
we perform a separate stellar calculation, imposing as outer boundary
conditions those appropriate for the conditions at the position of the
secondary inside the common envelope and taking into account the
contribution to gravity from the primary.

We obtain the angular-velocity profile in the envelope by solving the
equations of motion for the angular velocity, produced only by angular
momentum transport. These equations are averaged over azimuthal
and polar angles. The viscosity is assumed to be produced by turbulent
convection in convective regions (Meyer \& Meyer-Hofmeister 1979) and
to be the ordinary viscosity (i.e., molecular plus radiative) in
radiative regions.  The equations are solved with the explicit
conservation of the total angular momentum $J$ of the binary system,
simultaneously with the evolution of the common envelope (the stellar
structure is updated at each time step where we use the frictional
energy as determined from the current angular-velocity profile).  It
is further assumed that the primary star initially has a small angular
velocity compared to the orbital angular velocity.

\section{Stream-Core Interaction}

\subsection{Conditions at {\it L}$_1$ and the Ballistic Stage}

When the secondary starts to fill its own Roche lobe
inside the common envelope, we can estimate the mass-transfer
rate, $\dot{M}_{\rm sec}$, assumed to be adiabatic, according to 
(Ritter 1996)

\begin{displaymath} 
{\frac {\dot M_{\mbox{\scriptsize sec}}} 
{M_{\mbox{\scriptsize sec}}} } 
=  {\frac {1} {\zeta_{\rm ad}-\zeta_R}} 
{\frac {\partial \ln A} {\partial t} }
\end{displaymath} 

\noindent  where $M_{\rm sec}$ is the mass of the secondary, $A$  the  
binary separation,  $\zeta_{\rm ad}$  the
adiabatic  mass-radius exponent  and  $\zeta_R$  the mass-radius
exponent of the Roche radius. This estimate leads to characteristic
mass-loss rates of $10\,$--$\,100\,M_{\odot}\,
{\mbox{yr}}^{-1}$ for the binary parameters considered.

For the stream  cross-section  we  use 

\begin{displaymath} 
S_{\mbox{\scriptsize  CS}} = 1.5 \cdot 10^{20}\, s_{15}^{1/2} 
\Omega_{-3}^{-4/3} \dot M_{\mbox{\scriptsize  sec }} ^{-1/3}
\,\,{\rm cm}^2,
\end{displaymath}

\noindent where $s_{15} = {\frac {P} {\rho^{\gamma}}} \cdot 10^{-15}$,
$\Omega_{-3}=\Omega \cdot 10^3$ (both in cgs units), and $\dot M_{\rm sec}$ is 
the mass-loss rate in solar masses per year.

The entropy of the stream material is taken to be the same as the
entropy in the secondary.  Following the results of 3-dimensional 
numerical simulations for the stream at the inner Lagrange Point
(Bisikalo et al.\ 1998), we take the density profile of the stream
as being composed of a dense core with exponentially decreasing density
outside the core.  The velocity at $L_1$ is taken to be the local sound
speed of the stream material at $L_1$ (this speed varies through
the stream cross-section).

Initially, we take the stream trajectory to be ballistic up to the
point where the density of the ambient matter (the matter in the
common envelope) can no longer be ignored. The stream parameters at
this point are determined from the assumption that the Bernoulli
integral is conserved.  The angle, $\Theta$, at which the stream leaves $L_1$
is calculated using the formulae from Lubow \& Shu (1975).  This
approximation works well for the co-rotating frame and helps to avoid
numerical problems at $L_1$ at the start of the hydrodynamical
calculations.  Some uncertainty is introduced by the possibility that
the inner region surrounding the primary core is not in complete
co-rotation with the orbit and will have a backward motion with respect
to the co-rotating frame.  This relative rotation causes a force which
reduces the interception angle between the falling stream and the
normal direction to the core as compared to what a ballistic
trajectory would predict.

\subsection{The Hydrodynamical Stage}

For the hydrodynamical simulations of the stream-core interactions, we
use the hydrodynamical PROMETHEUS code (Fryxell, M\"{u}ller \& Arnett
1989 ) with the following modifications: an optional nuclear reactions
network (a full network including several hundred heavy elements and a
smaller network for the reactions of the hot CNO-cycle) and the
inclusion of the external gravitational field.  We perform 2-dimensional
calculations (using a cylindrical polar coordinate system).  The
common envelope of the star is treated as a stratified ambient medium
with a power-law dependence for the temperature and the pressure
distribution: $ T(r)=T(r_0) \left ( {\frac {r_0} {r}}\right
)^{\alpha_{T}} $ and $P(r)=P(r_0) \left ( {\frac {r_0} {r}}\right
)^{\alpha_{P}},$ where $\alpha_{T}=0.9-1.5$ and $\alpha_{P}=4.2-5.5$
well describe stellar models at this evolutionary stage in the region
we are interested in.

The boundary conditions which we use to describe the stream inflow are
taken to be the Mach number of the gas inflow, $M_{\mbox{\scriptsize
str}}$, the Mach number with respect to the ambient matter,
$M_{\mbox{\scriptsize amb}}$ (the ratio of the stream velocity to the
local ambient sonic speed), and the ratio between the central flow
density and the ambient density at the top of the box, $\eta_{\rho}$.
Radial and azimuthal velocities of the stream are defined by the
interception angle $\Theta$.  For the chemical composition of the gas
inflow we use the abundances obtained directly from the secondary
star. Boundary conditions for the rest of the box in the radial
direction are outflow boundaries (except in the gas inflow zone) with
the condition of hydrostatic equilibrium imposed, and in the azimuthal
direction are outflow conditions for non-symmetrical cases.
Alternatively one of the boundaries is taken to be a reflection boundary
(for symmetrical cases with $\Theta = 0$).

\section{Results}

\begin{figure}
\plotfiddle{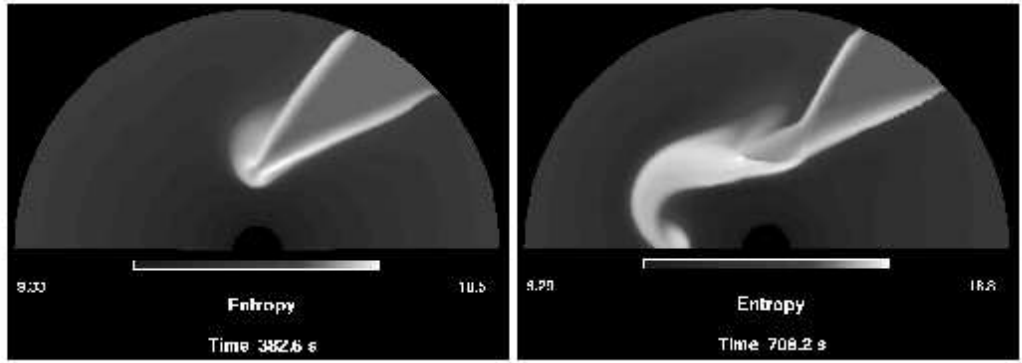}{5.5cm}{0}{75}{75}{-250}{-430}
\caption{Examples of the entropy  distribution for the stream-core interaction
at different times ($M_{\mbox{\scriptsize   str  }}   =  5.0$,
$M_{\mbox{\scriptsize amb }} = 2.5$,  $\eta_{\rho}= 5$, $\alpha_{ T } =
0.9$,  $\alpha_P  =  4.2$,   $\rho_{\mbox{\scriptsize  amb  }}  =  0.7
{\rm g}/\,{\rm cm}^{-3}$,
$T_{\mbox{\scriptsize amb }} = 2.5 \cdot 10^7\,$K). }
\end{figure}

We have modeled the common-envelope phase for systems composed of
$20+5, 20+1, 15+5\, M_\odot$ components, respectively, taking into
account the accretion luminosity of the secondary, the frictional
luminosity and the effect of the gravity from the secondary.
Parametrized results of these calculations were then used as initial
conditions and as boundary conditions for the hydrodynamical
calculations.  Together with calculations of the ballistic trajectories
and the Bernoulli integrals we estimated the parameter space of Mach
numbers typically found in our problem.  We then performed hydrodynamical
calculations for a number of parameter combinations
$(M_{\mbox{\scriptsize str }}, M_{\mbox{\scriptsize amb }})$. In
addition, for some Mach numbers, we performed calculations which
included nuclear burning. Purely hydrodynamical calculations showed
that the stream is able to penetrate deep into the dense region of the
primary for the assumed mass-loss rates and stream material
entropies. On the other hand, when nuclear burning is included, this
leads to the fast expansion of these streams for high $\dot M$ gas
inflow.  In this case, at least in 2-dimensional calculations,
hydrogen-rich material does not penetrate deep enough to start a hot
CNO cycle.  Instead, there is effective mixing in an intermediate
zone, which can lead to the dredge-up of He. In the case of gas inflow
with lower $\dot M$  and lower entropy material (e.g., as would be found 
for a $1 M_\odot$ secondary), a colder stream can penetrate deeper before the
temperature in the hydrogen-rich stream increases. Therefore in this
case the hot-CNO cycle with corresponding neutron production and
subsequent s-process may be possible.

\end{document}